\documentstyle[amssymb,preprint,aps]{revtex}

\begin{document}
\title{Fission of Multielectron Bubbles in Liquid Helium}
\author{J. Tempere$^{1,2}$, I.\ F. Silvera$^{1}$, J. T. Devreese$^{2}$}
\address{$^{1}$ Lyman Laboratory of Physics, Harvard University, Cambridge MA02138. \\
$^{2}$ Theoretische Fysica van de Vaste Stoffen, Universiteit Antwerpen,
B2610 Antwerpen, Belgium. }
\date{August 8, 2002}
\maketitle

\begin{abstract}
The stability of multielectron bubbles (MEBs) in liquid helium is
investigated using the liquid-drop model for fissioning nuclei. Whereas a
critical positive pressure can make the bubble unstable against fissioning,
a small negative pressure suffices to introduce a restoring force preventing
any small deformation of the bubble to grow. We also find that there exists
an energy barrier making MEBs metastable against fissioning at zero
pressure. The results obtained here overcome the difficulties associated
with the Rayleigh-Plesset equation previously used to study bubble
stability, and shed new light on the limits of achievable bubble geometries
in recently proposed experiments devised to stabilize MEBs.
\end{abstract}


\tightenlines

\bigskip 

\section{Introduction}

Multielectron bubbles (MEBs) are fascinating entities which appear when a
surface of bulk helium covered by a 2D film of electrons becomes unstable 
\cite{VolodinJETP26,AlbrechtEPL3}. MEBs are cavities in the liquid helium,
filled with electrons that form a 2D spherical layer at the inner surface of
the bubble. Recent proposals to stabilize MEBs \cite{SilveraPROC} have
stimulated theoretical investigations into its properties \cite
{LenzPRL87,TemperePRL87}, since this system holds the promise of studying
the electron gas in a controlled way, bereft from material impurities. The
density of this electron gas can be tuned over more than four orders of
magnitude by pressurizing the liquid helium \cite{TemperePRL87} and this
tunability would make the observation of a hexatic phase \cite{LenzPRL87}
and the quantum melting of a ripplopolaron Wigner lattice \cite
{KliminPRLs,FratiniEPJB14} experimentally feasible. For these
investigations, both theoretical and experimental, the question of the
stability of multielectron bubbles is of crucial importance. In this paper,
the results reported in a recent letter \cite{TemperePRL87} on the pressure
dependence of the frequency of the modes of deformation of an MEB are
discussed in the framework of new results obtained using the Bohr model for
fissioning \cite{BohrPR56}.

The energy and the radius of a multielectron bubble with $N$ electrons can
be estimated by balancing the surface tension with the electrostatic Coulomb
repulsion \cite{ShikinJETP27}. In this approximation, the energy of the MEB
is proportional to $N^{4/3}$, so that the energy of a single bubble with $N$
electrons is larger than the energy of two bubbles at infinite distance and
with $N/2$ electrons each. The lower energy of the fissioned bubble has led
to speculations about the stability of multielectron bubbles. Since MEBs
with $N$ up to $10^{8}$ were observed experimentally \cite
{VolodinJETP26,AlbrechtEPL3}, albeit in a transient manner (lasting a few
msec), there must exist a formation barrier preventing the fissioning of
MEBs. Early investigations ruled out gravitationally induced instabilities
and tunneling decay of the bubbles as possible fissioning mechanisms \cite
{ShikinJETP27}. Salomaa and Williams \cite{SalomaaPRL47} considered
dynamical stability against ``boiling off'' a single electron from a
multielectron bubble and found stability against this type of fission for
bubbles with $N>15-20$.

Preceding studies of the small amplitude oscillations of the bubble shape 
\cite{SalomaaPRL47,TemperePRL87} have shown that the quadrupole mode of
oscillation has a vanishing frequency when no pressure is applied on the
liquid helium. Furthermore it was shown that with increasing positive
pressure successive modes can be driven to a vanishing frequency \cite
{TemperePRL87}. The amplitude of modes of deformation that have a vanishing
frequency can grow until they become of the order of the bubble radius. This
deformational instability can lead to fissioning of the bubble. Salomaa and
Williams \cite{SalomaaPRL47} have investigated the dynamics of this
deformational instability using coupled Rayleigh-Plesset equations for the
deformation amplitudes and the bubble radius, and found that when the
initial amplitude of the quadrupole mode of deformation is larger than $\sim
6\%$ of the bubble radius, the amplitude of the oscillation keeps growing as
a function of time. However, their method is not valid when the oscillation
amplitude becomes comparable to the radius of the bubble; hence we need to
develop a new approach to describe the fission process.

\bigskip

\section{The Bohr model for fissioning multielectron bubbles}

In this paper, we apply the Bohr liquid-drop model of fissioning nuclei \cite
{BohrPR56} to MEBs. This method is valid for both small and large amplitude
deformations of atomic nuclei and describes the nucleus as a charged droplet
with surface tension. It was recently successfully improved to derive the
fragment mass asymmetries in nuclear fission \cite{MollerNAT409}. In such
models of nuclear fission, and similar models for the breaking up of
homogeneously charged liquid droplets, an approximation is used to calculate
the properties of the fission process. This approximation consists of
describing the surface of the splitting nucleus (or droplet) in terms of two
or three quadratic forms (spheroids and hyperboloids). Such a surface is
parametrized in cylindrical coordinates through 
\begin{equation}
\rho =\left\{ 
\begin{array}{l}
\sqrt{a_{L}-b_{L}(z-f_{L})^{2}}\text{ for }z_{0}\leqslant z\leqslant z_{1}
\\ 
\sqrt{a_{M}-b_{M}(z-f_{M})^{2}}\text{ for }z_{1}\leqslant z\leqslant z_{2}
\\ 
\sqrt{a_{R}-b_{R}(z-f_{R})^{2}}\text{ for }z_{2}\leqslant z\leqslant z_{3}
\end{array}
\right.  \label{cylrad}
\end{equation}
where

\begin{description}
\item  $a_{i}$ is the square of the semimajor axis of the spheroid along the
radial direction. For a hyperboloid, $a_{i}$ is negative ($i=L,M,R$);

\item  $b_{i}$ is the deformation parameter: the square of radial semimajor
axis divided by the square of the longitudinal semimajor axis;

\item  $f_{i}$ is the centre of the spheroid (hyperboloid);

\item  $z_{0}=f_{L}-a_{L}/\sqrt{b_{L}}$ is the leftmost point of the shape;

\item  $z_{3}=f_{R}+a_{R}/\sqrt{b_{R}}$ is the rightmost point of the shape.
\end{description}

\noindent These parameters are illustrated in Fig. 1. The surface determined
by (\ref{cylrad}) describes the shape of the bubble and allows to
investigate both spheroidal bubbles and emerging spheroidal fragments. The
shape parameters $\{a_{L},a_{M},a_{R}\},$ $\{b_{L},b_{M},b_{R}\},$ $%
\{f_{L},f_{M},f_{R}\},$ $\{z_{1},z_{2}\}$ are not independent if one imposes
continuity and continuous derivatives at the meeting points of the quadratic
forms. These conditions, together with fixing the origin at $z_{0}$ to
remove translations from the set of shape changes under consideration,
allows to eliminate five of the eleven variables. The six independent
parameters that are kept in our treatment are: $%
a_{L},b_{L},a_{R},b_{R},f_{R}-f_{L}$ and $z_{1}$. For a droplet of
incompressible fluid, there would be another constraint (that of fixed
volume) to remove one more parameter, but in the case of MEBs the volume
does not have to remain constant during deformations \cite{TemperePRL87}.

Within this model, an expression for the energy of a bubble with a given
shape (fixed by choosing the shape parameters $%
a_{L},b_{L},a_{R},b_{R},f_{R}-f_{L}$ and $z_{1}$) is obtained. The stable
shape of the bubble is found by minimizing the energy as a function of the
shape parameters -- the shape parameters set up a `shape space'. Both the
case of the spherical, unsplit bubble and the case of a bubble split in two
fragments can be described with appropriate shape parameters, so these two
cases can be represented by distinct points in the shape space. The dynamics
of the fissioning of the multielectron bubble can then be studied by
determining the energy along the trajectories in shape space which go from
the point representing the spherical, unsplit bubble to the point
representing a bubble split in two. The unsplit bubble may have a higher
energy than the fissioned bubble, but if there is an energy barrier along
each trajectory in shape space, then the single bubble is metastable with
respect to fissioning. The height of the energy barrier along the optimal
fissioning trajectory is the metastability energy barrier.

In section III we set up the expression for the energy of a bubble with
given shape parameters (the energy of a given point in shape space). In
section IV, we discuss the results of the minimization of the energy in
shape space and the results for the optimal trajectory for fissioning of a
multielectron bubble.

\bigskip

\section{Energy of a deformed MEB}

The energy of an MEB is determined as a function of the shape of the bubble
by three contributions: (i) the surface tension energy $E_{\sigma }=\sigma S$
where $\sigma =3.6\times 10^{-4}$ J/m$^{2}$ is the surface tension of liquid
helium and $S$ is the bubble surface, (ii) the pressure-related energy $%
E_{p}=pV$ where $p$ is the (experimentally tunable) difference in pressure
inside and outside the bubble and $V$ is the volume of the MEB; and (iii)
the electrostatic repulsion energy $E_{C}$ of the electrons. The first two
terms are easily evaluated since the surface and volume of the bubble are
related straightforwardly to the shape parameters. The surface is given by 
\begin{equation}
S=\sum_{i=1,2,3\equiv L,M,R}\text{ }\pi 
\displaystyle\int %
\limits_{z_{i-1}-f_{i}}^{z_{i}-f_{i}}\sqrt{a_{i}+b_{i}(b_{i}-1)x^{2}}dx,
\label{surf}
\end{equation}
where in the case of split bubbles ($b_{M}<0$ and $a_{M}<0$) the integration
domain should not include the region of space in between the bubble
fragments. The volume is 
\begin{equation}
V=%
\displaystyle\int %
\limits_{f_{L}-\sqrt{a_{L}/b_{L}}}^{f_{R}+\sqrt{a_{R}/b_{R}}}dz\text{ }\pi
\rho ^{2}(z),  \label{vol}
\end{equation}
where the cylindrical radius $\rho (z)$ is given by (\ref{cylrad}). The
integral in expression (\ref{vol}) is a piecewise sum of integrals of the
type 
\begin{equation}
\displaystyle\int %
\limits_{z_{a}}^{z_{b}}\pi \left[ a+b(z-f)^{2}\right] dz=\pi a(z_{b}-z_{a})-%
\frac{b}{3}((z_{b}-f)^{3}-(z_{a}-f)^{3}).
\end{equation}
The evaluation of the electrostatic energy is greatly simplified by the
observation \cite{ShungPRB45} that the quantum mechanical correction (the
exchange term) is negligible for the determination of the total
electrostatic energy. Furthermore the electrons in the bubble are not
smeared out throughout the bubble volume, but remain in a nanometer thin,
effectively two-dimensional layer anchored to the helium surface \cite
{ShikinJETP27,SalomaaPRL47}. This layer conforms to the helium surface also
when the bubble deforms \cite{SalomaaPRL47}. We will assume that the surface
density of electrons is homogeneous along the surface and equal to $n_{\text{%
s}}=N/S$ where $N$ is the number of electrons and $S$ is the area of the
deformed bubble. Some justification of this assumption comes from a recent
calculation of the coupled ripplon-phonon modes of oscillation of an MEB 
\cite{KliminPRLs}. In the calculation of Ref. \cite{KliminPRLs}, the
coupling between the modes of deformation of the helium surface
(``ripplons'') and the redistribution of the surface density of electrons
(``phonons'') was investigated, and it was derived that this coupling was
weak enough not to affect strongly the oscillation frequencies of ripplons
and phonons. Within the assumptions described in this paragraph, we may
write the electrostatic repulsion energy as 
\begin{eqnarray}
E_{C} &=&\int d^{3}{\bf r}\int d^{3}{\bf r}^{\prime }\frac{(n_{\text{s}%
}e).(n_{\text{s}}e)}{\varepsilon |{\bf r}-{\bf r}^{\prime }|}  \label{Ec} \\
&=&\frac{(2\pi n_{s}e)^{2}}{\varepsilon }%
\displaystyle\int %
\limits_{z_{0}}^{z_{3}}dz%
\displaystyle\int %
\limits_{z_{0}}^{z_{3}}dz^{\prime }\frac{2}{\pi }%
\displaystyle\int %
\limits_{0}^{\infty }dk\cos [k(z-z^{\prime })]I_{0}(k\rho _{<})K_{0}(k\rho
_{>})\rho (z)\rho (z^{\prime }),
\end{eqnarray}
where $\rho _{<}=\min [\rho (z),\rho (z^{\prime })]$ is the smallest of the
two cylindrical radii and $\rho _{>}=\max [\rho (z),\rho (z^{\prime })]$ is
the largest. $I_{0}$ is the modified Bessel function of the first kind with
index $0$, and $K_{0}$ is the modified Bessel function of the second kind\
with index $0$.

The total energy of a deformed bubble with given shape parameters is then
given by 
\begin{equation}
E=\sigma S+pV+E_{C}.  \label{energy}
\end{equation}
Expressions (\ref{surf}),(\ref{vol}), and (\ref{Ec}) allow to calculate the
total energy $E$ of the bubble for any point in the shape space discussed in
the previous section. In Ref. \cite{TemperePRL87} the energy of an MEB
undergoing small-amplitude oscillations was calculated. The total energy (%
\ref{energy}) of the deformed bubble used in the present treatment agrees
perfectly with the results of Ref. \cite{TemperePRL87} in those cases where
both approaches are applicable. Also the equilibrium radius of a spherical
bubble, obtained by minimizing the expression (\ref{energy}) with respect to
the radius, obviously agrees with the result of \cite{TemperePRL87}.

\bigskip

\section{Results and discussions}

To describe the fissioning of the bubble through a shape deformation, we
will investigate the trajectory in shape space that has the lowest energy
and that starts from the spherical bubble with equilibrium diameter $%
d=z_{3}-z_{0}=2R_{\text{b}}$. The equilibrium radius $R_{\text{b}}$ for the
spherical bubble with $N$ electrons can be found by minimizing 
\begin{equation}
E^{\text{spherical}}=4\pi \sigma R_{\text{b}}^{2}+%
{\displaystyle{4\pi  \over 3}}%
pR_{\text{b}}^{3}+%
{\displaystyle{N^{2}e^{2} \over 2\varepsilon R_{\text{b}}}}%
.  \label{Espherical}
\end{equation}
When $p=0,$ this radius is $R_{\text{b}}=\sqrt[3]{N^{2}e^{2}/(8\pi
\varepsilon \sigma )}$ (for example, with $N=10^{4}$ electrons in the
bubble, $R_{\text{b}}=1.064$ $\mu $m). To investigate the presence of an
energy barrier stabilizing the MEB against fissioning, we calculate the
minimum energy (and shape) of a bubble as a function of $d=z_{3}-z_{0}$, the
elongation of the shape along the axis of symmetry (see Fig. 1).

\subsection{Zero pressure}

The results of this minimization are shown in Fig. 2, for an MEB with $%
N=10^{4}$ electrons at $p=0$. For every given value of $d$ (the x-axis), the
shape parameters $a_{L},b_{L},a_{R},b_{R},f_{R}-f_{L},z_{1}$ are varied
under the constraint $z_{3}-z_{0}=d$. The optimal variational energy per
electron $E/N$ given by expression (\ref{energy}) is shown in Fig. 2, along
with the optimal variational shape for some selected points (marked by
A,B,E).

At $d=2R_{\text{b}}$, we find that the optimal shape is a spherical bubble
(the point marked by A in Fig. 1). At increased $d$, the optimal bubble
shape becomes ellipsoidal (for example, point B). The energy as a function
of $d$, for $d<3.012$ $\mu $m, is independent of $d$. This is in agreement
with previous results on the frequency of the vibrational modes of the
bubble \cite{SalomaaPRL47,TemperePRL87}. The eigenmodes of vibration of the
bubble surface are characterized by spherical harmonic mode numbers $\{\ell
,m\}$, and the eigenfrequencies (at $p=0$) are given by \cite
{SalomaaPRL47,TemperePRL87}: 
\begin{equation}
\omega (\ell )=\sqrt{\frac{\sigma }{\rho R_{\text{b}}^{3}}(\ell -2)(\ell
^{2}-1)}  \label{freqs}
\end{equation}
The deformation corresponding to point B in Fig. 2 is an $\ell =2$ eigenmode
of the system. This mode has zero frequency according to expression (\ref
{freqs}). This means that it does not cost energy to introduce small
amplitude $\ell =2$ deformations of the bubble, and the energy should remain
constant as a function of $d$, as it does (see Fig. 2).

The curve representing the minimum variational energy as a function of $d$
has a sudden change of slope near $d=3.012$ $\mu $m. The inset shows more
results near this point. We found that near this point two different minima
exist, corresponding to two topologically distinct shapes. On the horizontal
part of the curve (dashed line containing point B or point C in the inset of
Fig. 2) the shape is an ellipsoid. On the curve with negative slope (full
curve containing point E or point D in the inset of Fig. 2), the optimal
shape is a bubble split up into two fragments with $N/2$\ electrons in each
of the fission fragments. These two topologies compete for the global
minimum. For $d>3.012$\ $\mu $m, the fissioned bubble has lower energy:
point D is lower in energy than point C.

Let's investigate whether there exists some energy barrier stabilizing the
elongated bubble of point C against fission towards the fissioned shape
corresponding to D. To split up an elongated bubble (C) into fission
fragments (D), the bubble shape has to deform through intermediate shapes,
as those shown in Fig. 3. These intermediate shapes appearing during the
fission trace out a trajectory in shape space, starting at the point
corresponding to C and ending in the point corresponding to D. This
trajectory can be parametrized by an interpolation parameter which is zero
at the starting point (C) and 1 at the endpoint (D). The energy of the
intermediate shape as a function of the interpolation parameter is shown in
Fig. 3. It is clear that the two minima (elongated bubble, interpolation
parameter=0, and split-up bubble, interpolation parameter=1) are separated
by an energy barrier, of the order of 0.2 eV per electron. During the
process of fission, the elongated bubble has to be deformed to create the
neck between the fission fragment and the parent bubble. This deformation
must involve high-$\ell $\ modes of deformations, which cost energy $\omega
(\ell )>0$. This gives rise to the energy barrier shown in Fig. 3.

At this point it is also possible to clarify why the main limitation of the
Bohr model of fissioning does not affect our result. This limitation is that
only the splitting off of a single fragment -albeit of any size- can be
described. A fissioning process whereby the bubble splits in three or more
parts cannot be modelled realistically using only three quadratic surfaces.
Nevertheless it is clear that such a fissioning process would involve high-$%
\ell $ modes of deformation: the more fragments appear at the surface, the
more complicatedly it has to be deformed. Such a fissioning process would
therefore require even higher energy and its realization would be much less
likely. Note that the size of the fission fragments ($N/2$) is in agreement
with expression (\ref{Espherical}) for the energy of a spherical bubble, $E^{%
\text{spherical}}(N)\propto N^{4/3}$. This can be easily shown from
minimizing $E^{\text{spherical}}(m)+E^{\text{spherical}}(N-m)$\ with respect
to the optimal fragment size $m$.

\subsection{Effect of positive pressure}

In Ref. \cite{TemperePRL87}, it was shown that with increasing pressure,
more and more eigenmodes of the bubble deformation obtain vanishing
frequencies. The pressure dependence of the frequency of these modes is
given by \cite{TemperePRL87}: 
\begin{equation}
\omega (\ell )=\sqrt{\frac{\ell +1}{\rho R_{\text{b}}^{3}}\left[ \sigma
(\ell ^{2}+\ell +2)+2pR_{\text{b}}-\frac{N^{2}e^{2}}{4\pi \varepsilon R_{%
\text{b}}^{3}}\ell \right] }  \label{freqpospres}
\end{equation}
where $R_{\text{b}}$ is now the equilibrium radius under pressure, which
satisfies $2pR_{\text{b}}+4\sigma =e^{2}N^{2}/(4\pi \varepsilon R_{\text{b}%
}^{3})$. The effect of positive pressure is such that some modes of
deformation become `soft' (i.e. have vanishing frequency): for example, for
a bubble with $10^{4}$\ electrons at $p=3$\ mbar ($R_{\text{b}}=0.95$\ $\mu $%
m) the $\ell =2$\ and $\ell =3$\ modes of deformation are unstable \cite
{TemperePRL87}.

Fig. 4 shows the result for the minimum variational energy of a bubble with $%
10^{4}$ electrons at $p=3$ mbar ($R_{\text{b}}=0.95$ $\mu $m) as a function
of the elongation $d$. As the bubble deforms (along the curve containing
points ABC in Fig. 4), its energy decreases. Also, in contrast with the
zero-pressure case, the lowest energy shape for $d<2.68$\ $\mu $m exhibits
deformations with both $\ell =2$\ and $\ell =3$\ character. This is in
agreement with (\ref{freqpospres}).

Again two bubble topologies compete for the global minimum: the solution
where the shape is a singly connected surface (squares in Fig. 4), and the
solution where the bubble is split in two fragments (circles in Fig. 4). For 
$d<2.68$ $\mu $m the singly connected, unsplit bubble has the lower energy,
whereas at larger $d$, the fissioned bubble has lower energy.

The availability of higher angular momentum modes (such as $\ell =3$) allows
the MEB to deform to create a `neck' connecting an emergent fission fragment
with the `parent' bubble, without increasing the total energy of the bubble.
In the inset of Fig. 4, we show the variational energy of the intermediate
shapes assumed by the bubble during a fission process going from point C to
point D in Fig. 4. The `interpolation parameter' traces out the trajectory
in shape space during the fission process, as described in the discussion of
Fig. 2 in the previous subsection. The energy barrier which was present in
the case of zero pressure is no longer present, and we conclude that
pressurized MEBs are unstable when the pressure is high enough to drive $%
\ell >2$ modes unstable.

In ref. \cite{TemperePRL87} it was shown that the pressure at which a
particular mode of deformation becomes unstable, is larger for bubbles with
fewer electrons. Thus, when the pressure is raised so as to make a bubble
with $N$\ electrons unstable and fission that bubble, the resulting fission
products with $N/2$\ electrons may still be metastable. To fission also
those fragments the pressure needs to be raised further.

\subsection{Effect of negative pressure}

Liquid helium can sustain a negative pressure of $-9$ bar ($\pm 1$ bar)
before it cavitates \cite{MarisPRL63}. Here, a negative pressure on the MEB
means that the force associated with this pressure is directed outward, away
from the bubble center. Shikin \cite{ShikinJETP27} estimated in a simplified
model that MEBs with a radius larger than the zero-pressure equilibrium
radius $R_{\text{b}}(p=0)$ may be metastable in the sense that there exists
a restoring force which counteracts small deformations. Salomaa and Williams 
\cite{SalomaaPRL47} found that the nonlinear coupling between the radial
mode of oscillation and the deformational modes of oscillation may lead to a
small increase in the time-averaged radius, stabilizing the bubble. A
straightforward way to increase the bubble radius is by applying negative
pressure. In Ref. \cite{TemperePRL87} we showed that for negative pressures
all the eigenmodes of deformation acquire a positive frequency.

The liquid-drop model of fission allows to describe deformations beyond the
small-amplitude region of validity of the Rayleigh-Plesset equations used in 
\cite{SalomaaPRL47}. Figure 5 shows the result of negative pressures on the
shape and fissioning of the bubble. The variational minimum energy is again
shown as a function of the elongation along the axis of symmetry $%
d=z_{3}-z_{0}$, for a bubble with $N=10^{4}$ electrons at $p=-3$ mbar ($%
R_{b}=1.5$ $\mu $m). In contrast to unpressurized MEBs and MEBs at positive
pressure, now three geometries compete for the global minimum instead of
two: (A) an $\ell =2$ deformed bubble, (B) the split-up bubble, and (C) a
spherical bubble with expanded radius. To deform the bubble from any of
these three shapes into another, intermediate shapes have to be assumed and
an energy barrier will be present like in the case of zero pressure.

But, apart from this energy barrier, another energy barrier is present: from
the increase of the energy with increasing $d$\ in Fig. 5, it is clear that
there is a restoring force for small-amplitude deformations. This restoring
force was absent for $p=0$ (compare Fig. 5 with Fig. 2) and in the case of $%
p>0$ the force was of opposite sign and driving the instability (compare
Fig. 5 with Fig. 4). The restoring force is related to the lowest frequency
of the eigenmodes, $\omega (\ell =2)$ which becomes positive at negative
pressures, as derived in \cite{TemperePRL87}. The restoring force results in
an additional energy barrier of 0.15 eV per electron in the case studied
here. We emphasize that the additional energy barrier related to the
intermediate bubble shapes (as illustrated by Fig. 3 for the $p=0$ case) is
present also at negative pressures (but not at positive pressures large
enough to drive $\ell >2$ modes to zero frequency).

Note furthermore that, if there is a driving force which can excite the
deformation to large amplitude and overcome the barriers, the MEB may be
destroyed in one of two ways: either by fissioning of the bubble, after
which the fission fragments move away from each other, or the MEB may keep
expanding until it fills a volume large enough to counteract the negative
pressure (i.e. the MEB serves as a nucleation center for cavitation).

\section{Conclusions}

In this paper, the liquid drop model of fissioning was applied to the
problem of multielectron bubbles in liquid helium. We found that, even
though there exists, at $p=0$, a mode of deformation which can grow without
increasing the total energy of the bubble, there is still an energy barrier
present which prevents fissioning of the bubble. This barrier was explained
by the intermediate shapes that the fissioning bubble has to assume in order
to create a neck between the emerging fission fragment and the parent
bubble. These intermediate shapes involve eigenmodes of deformation which
cost substantial energy (0.2 eV per electron for a 10000 electron bubble).
At positive pressure, these higher angular momentum modes can obtain a
vanishing frequency as discussed in \cite{TemperePRL87}, and this causes the
energy barrier to vanish and the bubble to become unstable. However, at
negative pressure, when all the modes of deformation have a non-vanishing
frequency, there is an additional element of stability in that there is a
restoring force which counteracts small amplitude deformations. The present
study, based on the liquid drop model, independently confirms and extends
the conclusions presented in a previous letter \cite{TemperePRL87}, namely
that a positive pressure can make the MEB unstable, whereas a small negative
pressure makes the MEB metastable against fission. The additional result
presented in this paper is that also at zero pressure the MEB is metastable.

\section{Acknowledgments}

Discussions with J. Huang are gratefully acknowledged. J. Tempere is
supported financially by the FWO-Flanders with a mandate `Postdoctoral
Fellow of the Fund for Scientific Research - Flanders' (Postdoctoraal
Onderzoeker van het Fonds voor Wetenschappelijk Onderzoek - Vlaanderen).
This research has been supported by the Department of Energy, grant
DE-FG002-85ER45190, and by the\ GOA BOF UA 2000, IUAP, the FWO-V projects
Nos. G.0071.98, G.0306.00, G.0274.01, WOG WO.025.99N (Belgium).

\bigskip

\bigskip

\section*{Figure captions}

\bigskip 

{\bf NOTE: due to the size requirements of the figure files in the Los
Alamos preprint archive, the quality of the figures has been reduced in this
preprint - contact the authors for high-quality figures.}

\bigskip 

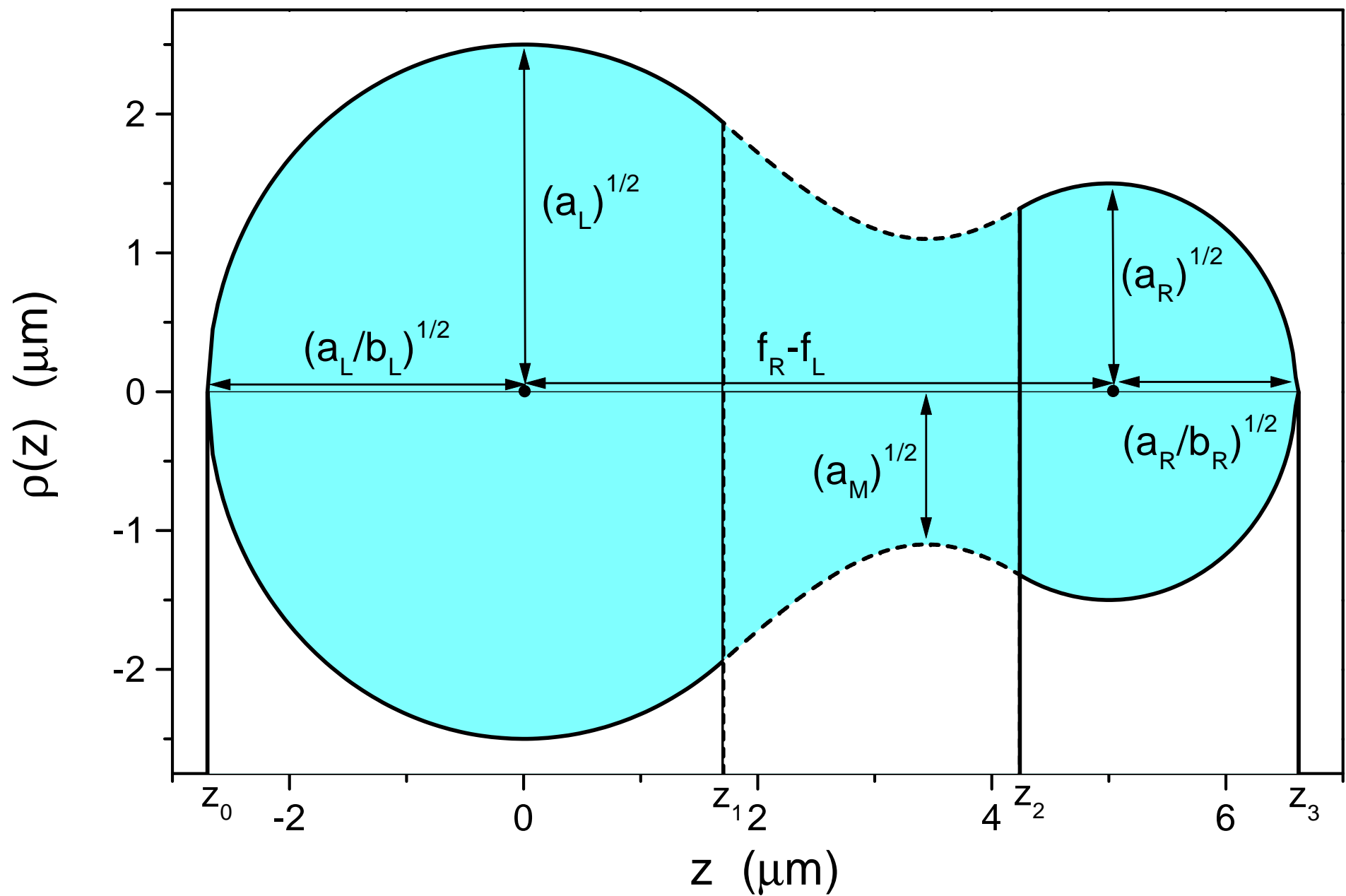
FIG. 1. The model shape for a fissioning MEB consists of three quadratic
forms, for example two spheroids connected by a hyperboloidal neck as
illustrated in this figure. The relation between the model shape and the
parameters used in the text, expression (\ref{cylrad}), is shown in this
figure.

\bigskip

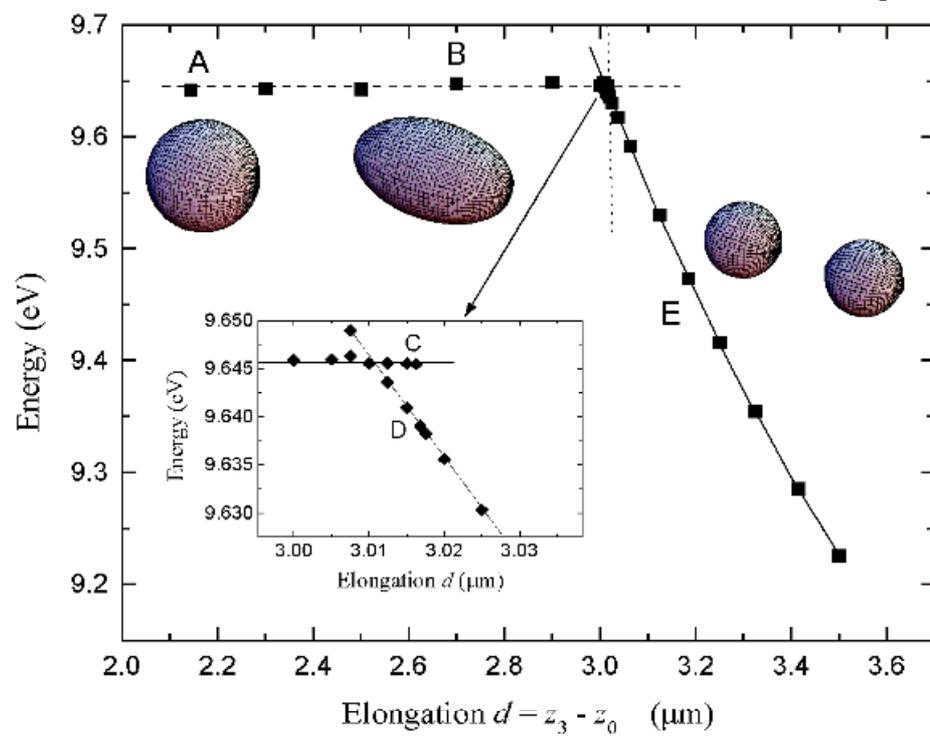
FIG. 2. The minimum variational energy per electron (in eV) of an $N=10^{4}$
electron MEB at $p=0$ is shown as a function of the bubble elongation $d$
(microns). This energy is obtained by minimizing expression (\ref{energy})
with respect to the shape parameters illustrated in Fig. 1, and subject to
the constraint $z_{3}-z_{0}=d$. The symbols (squares and diamonds) are the
results of the minimization and the curves are guides for the eye. For some
points of interest (A,B,E) the corresponding shape of the bubble is
illustrated. Two bubble shapes compete for the minimum energy: an
elliptically elongated bubble (points on the dashed curve) and the fissioned
bubble (points on the full curve). For $d<3.012$ $\mu $m the elongated
bubble is the minimum energy shape whereas for larger $d$ the fissioned
bubble is the minimum energy shape.

\bigskip

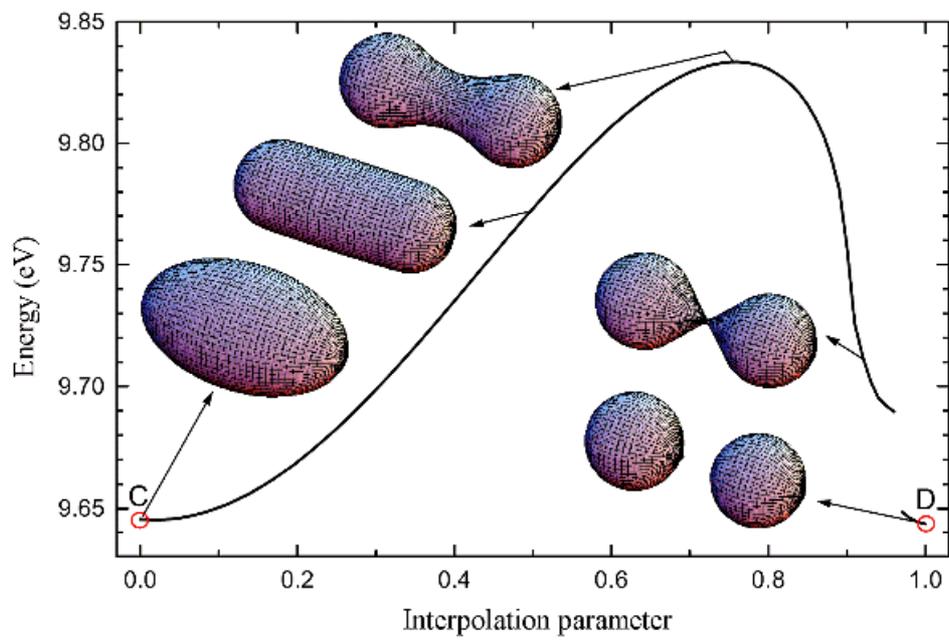
FIG. 3. When a $10^{4}$ electron MEB at zero pressure is elongated more than 
$d=3.012$ $\mu $m, the minimization of the total energy shows that it
becomes energetically favorable to split the bubble in two equal fragments.
However, in order to fission the elongated bubble, it has to deform in such
a way that a neck develops where the bubble can split in two. These
intermediate shapes of the bubble are higher in energy than either the
elongated shape (C) or the split bubble (D), and give rise to the presence
of an energy barrier. Points C and D in this figure correspond to those of
Fig. 2; the fission process traces out a trajectory in shape space
connecting C and D and parametrized by an interpolation parameter ranging
from 0 (C) to 1 (D).

\bigskip

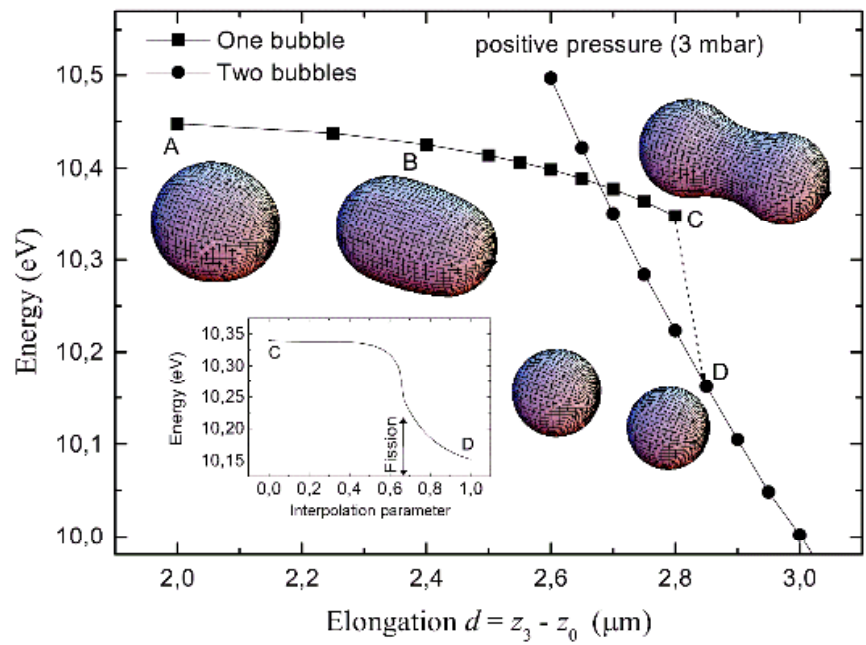
FIG. 4. The minimum variational energy per electron of an $N=10^{4}$
electron MEB at $p=3$ mbar is shown as a function of the bubble elongation $%
d $ (in micron). For some points of interest (A,B,C,D), the bubble shape is
illustrated. In the inset, the energy is shown as a function of the
interpolation parameter which takes the shape parameters from point C into
those from point D. Comparing this figure with Figs. 2,3, it is clear that
the bubble now can decrease its energy monotonically while splitting:
applying a positive pressure so that $\ell >2$ modes of deformation obtain a
vanishing frequency can remove the energy barrier and make the MEB\ unstable.

\bigskip

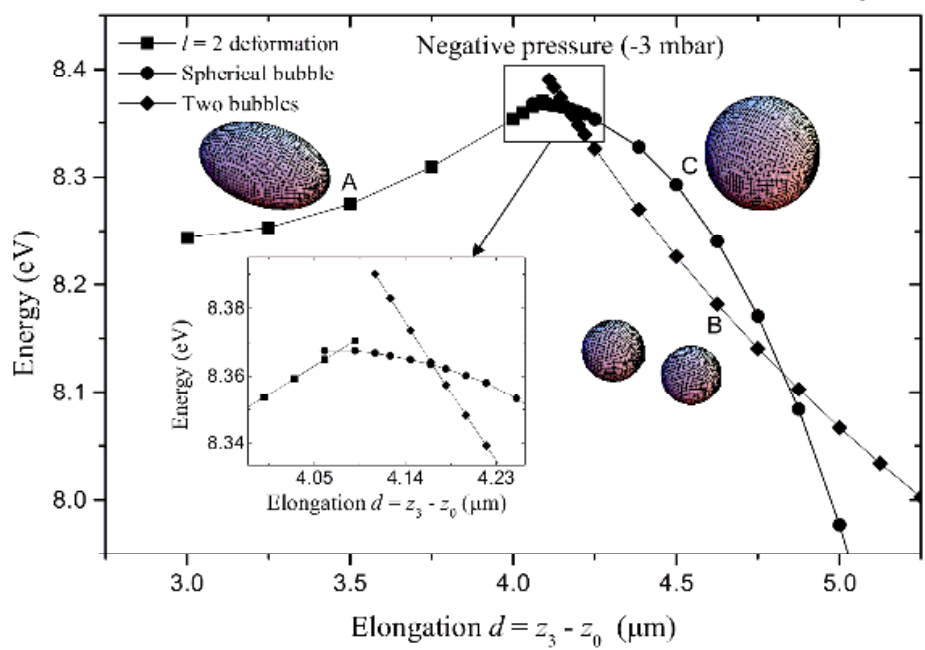
FIG. 5. The minimum variational energy per electron of an $N=10^{4}$
electron MEB at $p=-3$ mbar is shown as a function of the bubble elongation $%
d=z_{3}-z_{0}$ (in micron). Three geometries compete for the global minimum:
the $\ell =2$ mode of deformation of the bubble (point A), the split bubble
(point B) and the spherical bubble with large radius (point C). At negative
pressure, there exists an additional energy barrier associated with a
restoring force which prevents small-amplitude deformations ($d<4$ $\mu $m)
from growing. The inset shows the region near the maximum of the energy
barrier in more detail.

\end{document}